\documentclass[12pt]{article}

\ifx\pdfoutput\undefined
\usepackage[dvips,bookmarks]{hyperref}
\else
\usepackage{hyperref}
\fi
\hypersetup{colorlinks=false,bookmarksopen,bookmarksnumbered,citecolor=blue,
   pdfstartview=FitH}

\usepackage{latexsym}
\usepackage{amssymb,amsfonts,amsmath}
\usepackage{graphicx} 
\usepackage{indentfirst}
\usepackage{bbm}
\usepackage{amssymb}
\usepackage{verbatim}
\usepackage{amsmath, amsthm,amssymb}
\usepackage{mathrsfs}
\usepackage{hyperref}
\usepackage{amsfonts}
\usepackage{dsfont}
\usepackage{slashed, tensor}
\usepackage{booktabs}
\usepackage{graphicx}
\usepackage{mathrsfs}
\parskip=\medskipamount

\usepackage[mathscr]{euscript} % alternate script style

\newcommand {\cD}{{\cal D}}
\newcommand {\cE}{{\cal E}}

\newcommand {\cH}{{\cal H}}

\newcommand {\cL}{{\cal L}}
\newcommand {\cM}{{\cal M}}
\newcommand {\cN}{{\cal N}}
\newcommand {\cO}{{\cal O}}

\newcommand {\cR}{{\cal R}}

\newcommand {\cZ}{{\cal Z}}
\def\a{\alpha}
\def\b{\beta}

\def\d{\delta}
\def\e{\epsilon}
\def\f{\phi}
\def\g{\gamma}
\def\G{\Gamma}

\def\j{\psi}
\def\k{\kappa}
\def\l{\lambda}

\def\q{\theta}
\def\r{\rho}
\def\s{\sigma}

\def\x{\xi}
\def\z{\zeta}
\def\D{\Delta}

\def\J{\Psi}
\def\L{\Lambda}
\def\O{\Omega}

\def\U{\Upsilon}
\def\X{\Xi}
\def\ri{{\rm i}}
\def\re{{\rm e}}

\def\rA{{\rm A}}
\newcommand{\gd}{{\dot\g}}

\newcommand{\ad}{{\dot{\alpha}}}
\newcommand{\bd}{{\dot{\beta}}}

\newcommand{\sSU}{\mathsf{SU}}

\newcommand{\sU}{\mathsf{U}}

\newcommand{\1}{{\underline{1}}}
\newcommand{\2}{{\underline{2}}}

\newcommand{\ve}{\varepsilon}
\newcommand{\cDB}{{\bar\cD}}
\newcommand{\DB}{\bar{D}}

\newcommand{\pa}{\partial}
\newcommand{\hf}{\frac12}

\newcommand{\be}{\begin{equation}}
\newcommand{\ee}{\end{equation}}
\newcommand{\bea}{\begin{eqnarray}}
\newcommand{\eea}{\end{eqnarray}}
\newcommand{\non}{\nonumber}
\newcommand{\ba}{\begin{array}}
\newcommand{\ea}{\end{array}}

\newcommand{\bm}[1]{\mbox{\boldmath$#1$}}

\def\double #1{#1{\hbox{\kern-2pt $#1$}}}

\newcommand{\bsubeq}{\begin{subequations}}
\newcommand{\esubeq}{\end{subequations}}

\newcommand{\rd}{\mathrm d}
\newcommand{\de}{\nabla}

\numberwithin{equation}{section}  % Resets equation number at beginning of each section

\newcommand{\qb}{{\bar{\theta}}}

\renewcommand{\de}{{\nabla}}
\newcommand{\deb}{{\bar{{\nabla}}}}

\def\ft#1#2{{\textstyle{\frac{\scriptstyle #1}{\scriptstyle #2} } }}

\begin{document}
%%%%%%%%%%%%%%%%
%%%%%%%%%%%%%%%

\begin{titlepage}
\begin{flushright}
July, 2017 \\
\end{flushright}
\vspace{5mm}

\begin{center}
{\Large \bf 
New nilpotent \mbox{$\bm{\cN=2}$} superfields}
\end{center}

\begin{center}

{\bf
Sergei M. Kuzenko${}^{a}$ and
Gabriele Tartaglino-Mazzucchelli${}^{b}$
} \\
\vspace{5mm}

\footnotesize{
${}^{a}${\it School of Physics and Astrophysics M013, The University of Western Australia\\
35 Stirling Highway, Crawley W.A. 6009, Australia}}  
~\\
\vspace{2mm}
\footnotesize{
${}^{b}${\it Instituut voor Theoretische Fysica, KU Leuven,\\
Celestijnenlaan 200D, B-3001 Leuven, Belgium}
}
\vspace{2mm}
~\\
\texttt{sergei.kuzenko@uwa.edu.au,
gabriele.tartaglino-mazzucchelli@kuleuven.be}\\
\vspace{2mm}

\end{center}

\begin{abstract}
\baselineskip=14pt
We propose new off-shell models for spontaneously broken local $\cN=2$ 
supersymmetry, in which the supergravity multiplet couples
to nilpotent Goldstino superfields that contain either a gauge one-form or a gauge 
two-form in addition to  spin-1/2 Goldstone fermions and auxiliary fields. 
In the case of $\cN=2$ Poincar\'e supersymmetry, we elaborate on the concept of 
twisted chiral superfields and present a nilpotent $\cN=2$ superfield 
that underlies the cubic nilpotency conditions given in  arXiv:1707.03414
in terms of constrained $\cN=1$ superfields.
\end{abstract}

\vfill

\vfill
\end{titlepage}

\newpage
\renewcommand{\thefootnote}{\arabic{footnote}}
\setcounter{footnote}{0}

%\tableofcontents

%%%%%%%%%%%%%%%%%%%%%%%%%%%%%%%%%%%%%%%%%%%%%%%%%%%%%%
%%%%%%%%%%%%%%%%%%%%%%%%%%%%%%%%%%%%%%%%%%%%%%%%%%%%%%

\allowdisplaybreaks

\section{Introduction}

In the last three years, there has been much interest in 
off-shell models for spontaneously broken $\cN=1$ supergravity, see 
\cite{ADFS,DFKS,BFKVP,HY,K15,BMST,FKRR,BHKMS,BK17,KMcAT-M} 
and references therein.
Recently, we have constructed several off-shell models for spontaneously broken 
$\cN=2$ supergravity \cite{KMcAT-M}, in which the supergravity multiplet couples
to the nilpotent Goldstino superfields introduced in \cite{KMcA,CDF}. 
The models proposed in  \cite{KMcAT-M} make use of both reducible and irreducible Goldstino superfields, following 
the terminology of \cite{BHKMS}. 
Every irreducible $\cN=2$ Goldstino superfield contains only two independent
component fields -- the spin-1/2 Goldstone fermions \cite{VA,AV}, 
while the other component fields are composites constructed from the Goldstini.
Reducible Goldstino superfields also contain some independent fields 
in addition to the  Goldstini.

In this paper we propose new models for spontaneously broken 
$\cN=2$ supergravity in which the Goldstini belong to nilpotent superfields
containing either a gauge one-form or a gauge two-form among its independent physical fields. This will be achieved 
by relaxing the constraints obeyed
by the Goldstino superfields introduced in \cite{KMcAT-M,KT-M15}.
The idea can be illustrated by giving two examples. 
The oldest irreducible Goldstino superfield in four dimensions is 
the  $\cN=1$ chiral scalar superfield $\f$ 
\cite{Rocek,IK78},
$\bar D_{\dot\alpha} \f =0$, which is
subject to the constraints \cite{Rocek}: 
\begin{subequations} \label{1.1}
\bea 
\f^2&=&0 ~,  \label{1.1a}\\
{f} \f &=& -\frac 14  \f {\bar D}^2 \bar  \f ~, \label{1.1b}
\eea
\end{subequations}
where $f$ is a real parameter of mass dimension $+2$ which characterises 
the supersymmetry breaking scale. Removing the second constraint, eq. \eqref{1.1b},
leads to the reducible Goldstino superfield  advocated in 
\cite{Casalbuoni,KS}. Our second example is the irreducible Goldstino superfield
introduced in \cite{BHKMS}. It is described by a real scalar $\cN=1$ superfield
 $V$ subject to the nilpotency constraints\footnote{The constraints \eqref{1.2a} 
 and \eqref{1.6} were introduced for the first time in \cite{LR79}.}
\begin{subequations} \label{1.2}
\bea
V^2&=&0~, \label{1.2a}\\ 
V D_A D_B V &=&0~, \label{1.2b}\\
V D_A D_B D_C V &=&0~, \label{1.2c}
\eea
\end{subequations}
where $D_A =(\pa_a , D_\a, \bar D^\ad)$ are the covariant derivatives of 
$\cN=1$ Minkowski superspace,  in conjunction with the nonlinear constraint 
\bea
f  V =\frac{1}{16} V D^\a \bar D^2 D_\a V~.
\label{1.6}
\eea
If the nonlinear constraint \eqref{1.6} is removed, we end up with the reducible 
Goldstino superfield introduced in \cite{KMcAT-M}.

This paper is organised as follows. In section 2 we couple $\cN=2 $ supergravity 
to a deformed reduced chiral superfield subject to a cubic nilpotency condition. 
In section 3,  $\cN=2 $ supergravity is coupled 
to a linear superfield (also known as the $\cO(2)$ multiplet)  subject to a cubic nilpotency condition. In section 4, we 
elaborate on the concept of twisted chiral superfields in $\cN=2$ Minkowski superspace
and present a nilpotent $\cN=2$ superfield 
that underlies the cubic nilpotency conditions given in \cite{DFS}
in terms of constrained $\cN=1$ superfields. The reason for restricting our analysis  
to the super-Poincar\'e case is that there is no simple definition of twisted chiral superfields 
on arbitrary $\cN=2$ curved superspace backgrounds. 
The main body of the paper is accompanied by two technical appendices.
Appendix \ref{App1} reviews the prepotential formulations for  the $\cN=2$ 
reduced chiral and linear multiplets. Appendix \ref{App2}
provides a solution to the nilpotency condition \eqref{3.2}
in the flat case using the harmonic superspace techniques. 

%%%%%%%%%%%%%%%%%%%%%%%%%%%%
%%%%%%%%%%%%%%%%%%%%%%%%%%%%%

\section{Nilpotent chiral superfield}\label{section2}

In recent papers \cite{K-superWeyl,KT-M15},  a deformed 
reduced chiral superfield $\cZ$ coupled to $\cN=2$ supergravity was introduced. 
 It is described by the constraints
\begin{subequations}\label{Z}
\bea
\bar \cD^i_\ad \cZ &=&0~, \label{Za} \\
\big(\cD^{ij}+4S^{ij}\big)\cZ
- \big(\cDB^{ij}+ 4\bar{S}^{ij}\big)\bar{\cZ} &=&4  \ri G^{ij} ~, \label{Zb}
\eea
\end{subequations}
where we have defined $\cD^{ij}=\cD^{\a(i}\cD_\a^{j)}$
and $\cDB^{ij}=\cDB_\ad^{(i}\cDB^{j) \ad}$.
Here $G^{ij}$  is a linear multiplet which obeys the constraints \eqref{1.2}.
In addition, $G^{ij}$ is required to be nowhere vanishing,  $G^{ij} G_{ij} \neq 0$.
As reviewed in Appendix \ref{App1}, 
$G^{ij}$ is  the gauge-invariant field strength of a tensor multiplet. 
In this paper, we identify $G^{ij}$ with one of the two conformal compensators
of the minimal  formulation for $\cN=2$ supergravity proposed in \cite{deWPV}.
The superfields $S^{ij} $ and ${\bar S}^{ij} $  in \eqref{Z}
are special dimension-1 components of the torsion, see \cite{KLRT-M2} for the technical details of the superfield 
formulation for $\cN=2$ conformal supergravity \cite{Howe} that  we use. 
The constraints \eqref{Za} and \eqref{Zac} are invariant under the $\cN=2$ super-Weyl 
transformations \cite{KLRT-M2,Howe} if $\cZ$ is chosen to be a primary superfield of dimension 1.

In the super-Poincar\'e case, a chiral superfield obeying the constraint 
\eqref{Zb} with a {\it constant} $\sSU(2)$ triplet $G^{ij} $ appeared in the framework
of partial $\cN=2 \to \cN=1 $ supersymmetry breaking \cite{APT,IZ1,IZ2}.

In our previous paper \cite{KT-M15}, $\cZ$ was subject to the quadratic nilpotency 
condition\bea
\cZ^2 =0~. 
\label{Zac}
\eea
The constraints  \eqref{Z} and \eqref{Zac} imply that, for certain 
$\cN=2$ supergravity backgrounds,  
the degrees of freedom described by the $\cN=2$ chiral superfield $\cZ$ 
are in one-to-one correspondence with those of an Abelian $\cN=1$ vector multiplet. 
The specific feature of such $\cN=2$ supergravity backgrounds is that 
they possess an $\cN=1$ subspace $\cM^{4|4}$ of the 
full $\cN=2$ curved superspace $\cM^{4|8}$. 
This property is not universal. In particular, there exist maximally 
$\cN=2$ supersymmetric backgrounds with no admissible truncation to $\cN=1$ 
\cite{BIL}.
As shown in  \cite{KT-M15},
the superfield constrained by  \eqref{Z} and \eqref{Zac} 
is suitable for the description of
partial $\cN=2 \to \cN=1 $ rigid supersymmetry breaking 
in every maximally supersymmetric spacetimes $\cM^{4}$
which is the bosonic body of an  $\cN=1$ superspace $\cM^{4|4} $ described 
by the following algebra of $\cN=1$ covariant derivatives\footnote{These backgrounds
are maximally supersymmetric solutions of  pure $R^2$ supergravity
\cite{Kuzenko:2016nbu}.} 
\begin{subequations}
\label{RS^3}
\bea
&\{\cD_\a,\cD_\b\}= 0~, \qquad \{\cDB_\ad,\cDB_\bd\}=0~,\qquad
\{\cD_\a,\cDB_\bd\}=-2\ri\cD_{\a\bd}~,
\\
&{[}\cD_\a,\cD_{\b\bd}{]}=\ri\ve_{\a\b}G^\g{}_{\bd}\cD_\g
~,\qquad
{[}\cDB_\ad,\cD_{\b\bd}{]}=-\ri\ve_{\ad\bd}G_\b{}^\gd\cDB_\gd~,
\\
&{[}\cD_{\a\ad},\cD_{\b\bd}{]}=
-\ri\ve_{\ad\bd}G_\b{}^\gd\cD_{\a\gd}
+\ri\ve_{\a\b}G^\g{}_\bd\cD_{\g\ad}
~,
\eea
where the real four-vector $G_{b}$ is covariantly constant,
\bea
\cD_\a G_b = 0~, \qquad G_{b} =\bar G_b~.
\eea
\end{subequations}
Since $G^2 = G^b G_b $ is constant, the geometry 
\eqref{RS^3} describes  three different superspaces, for $G_b \neq 0$, 
which correspond to the choices $G^2<0$, $G^2>0$ and $G^2=0$, 
respectively. The Lorentzian manifolds $\cM^4$ supported by these superspaces are 
${\mathbb R}\times S^3$, ${\rm AdS}_3 \times S^1$ or its covering 
${\rm AdS}_3 \times {\mathbb R}$, 
and a pp-wave spacetime, respectively.
 
We constructed in \cite{KT-M15} the Maxwell-Goldstone multiplet actions
for partial $\cN=2 \to \cN=1$ supersymmetry breaking for all of them.
In each of these cases, 
the action coincides with a unique curved-superspace extension of the $\cN=1$ 
supersymmetric Born-Infeld action \cite{DP,CF,BG}, which is 
singled out by the requirement of $\sU(1)$ duality invariance \cite{KT2,KT3,KMcC}. 
In the super-Poincar\'e case, $G_b =0$, 
the approach developed in  \cite{KT-M15} provided a simple
$\cN=2$ superfield derivation of 
 the Bagger-Galperin action for partial $\cN=2 \to \cN=1$ supersymmetry breaking
\cite{BG}, which differs in some technical details from the original derivation 
given by Ro\v{c}ek and Tseytlin \cite{RT}.

If one is interested in $\cN=2 \to \cN=0 $ breaking of local supersymmetry, 
the nilpotency condition \eqref{Zac} should be replaced with a weaker constraint 
\bea
\cZ^3 =0~.
\label{2.4}
\eea
In the super-Poincar\'e case, such a constraint has recently been considered 
in \cite{DFS}. As was demonstrated  in \cite{DFS}, for a certain range of parameters, 
in Minkowski superspace the superfield $\cZ$  constrained by  \eqref{Z} and \eqref{2.4} 
contains the following independent fields:
two Goldstini,  a gauge one-form
and a real, nowhere vanishing, $\sSU(2) $ triplet of auxiliary fields $D^{ij} =D^{ji}$, with  
$D^{ij} D_{ij} \neq 0$. We now present a dynamical system describing $\cN=2$ supergravity coupled to $\cZ$.

The action for our supergravity-matter theory involves two contributions
\bea
S= S_{\rm SUGRA}  +S_\cZ~,
\label{2.5}
\eea
where $S_{\rm SUGRA}$ denotes the pure supergravity action 
and $S_\cZ$ corresponds to the goldstino superfield.
We make use of the minimal formulation for $\cN=2$ supergravity 
with vector and tensor 
compensators \cite{deWPV}.  
In the superspace setting, the supergravity action 
can be written in the form \cite{BK11}
(derived using the projective-superspace formulation \cite{K-08} for this theory)
\bea
S_{\rm SUGRA}  &=& \frac{1}{ \k^2} \int \rd^4 x {\rm d}^4\q \, \cE \, \Big\{
\J {\mathbb W} - \frac{1}{4} W^2 +m \J W \Big\}          +{\rm c.c.}   \non \\
 &=& \frac{1}{ \k^2} \int \rd^4 x {\rm d}^4\q \, \cE \, \Big\{
\J {\mathbb W} - \frac{1}{4} W^2 \Big\} +{\rm c.c.}
     + \frac{m}{ \k^2} \int \rd^4 x {\rm d}^4\q \rd^4\bar\theta\,
     E \, G^{ij}V_{ij}~,~~~~~~~
\label{6.11} 
\eea
where $\k$ is the gravitational constant and $m$ the cosmological parameter.
Here $E$ and $\cE$ denote the full superspace and chiral densities, respectively. 
The covariantly chiral scalar $\J$ and the real $\sSU(2)$ triplet $V_{ij}$ are the prepotentials of the tensor and vector 
multiplets, respectively, see Appendix \ref{App1} for the technical details.
The supergravity action involves the composite
\bea
   \mathbb W := -\frac{G}{8} (\bar \cD_{ij} + 4 \bar S_{ij}) \left(\frac{G^{ij}}{G^2} \right) ~, 
   \label{2.100}
\eea
which proves to be a reduced chiral superfield.
The superfield \eqref{2.100} is 
one of the simplest applications of the powerful approach 
to generate composite reduced chiral multiplets 
which was presented in \cite{BK11}. Another application will be given in 
the next section.

The action for the goldstino superfield $\cZ$ in \eqref{2.5} is 
\bea
S_\cZ  &=&  \int \rd^4 x {\rm d}^4\q \, \cE \, \Big\{
\frac{1}{4} \cZ^2  +\z W \cZ 
+\r \Big(\cZ \J - \frac{\ri}{2}  \J^2\Big)
\Big\}      +{\rm c.c.}   ~,
\label{2.7}
\eea
where $\z$ and $\r$ are  complex and real parameters, respectively.
The $\r$-term in \eqref{2.7} was introduced in  \cite{KT-M15},
where it was shown to be invariant under gauge transformations 
 \eqref{TMGT}.

The goldstino superfield action \eqref{2.7} can be generalised to include 
higher derivative couplings, for instance 
\bea
I=\int \rd^4 x {\rm d}^4\q \rd^4\bar\theta\,
     E \,\Big\{ \l_1 \frac{\bar \cZ \cZ}{\bar W W}
     +  \l_2  \Big(\frac{ \bar \cZ \cZ }{ \bar W W }\Big)^2 \Big\}~,
\eea
where $\l_1$ and $\l_2$ are coupling constants. 

%%%%%%%%%%%%%%%%%%%%%%%%%%
%%%%%%%%%%%%%%%%%%%%%%%%%%

\section{Nilpotent linear superfield}\label{section3}

We now introduce a linear superfield $\cH^{ij}$,
\bea
\cD^{(i}_\a \cH^{jk)} =  {\bar \cD}^{(i}_\ad \cH^{jk)} = 0~,
\label{3.1}
\eea
which is subject to the following cubic nilpotency condition \cite{KMcAT-M} 
\bea
{\cH}^{(i_1i_2} {\cH}^{i_3 i_4} {\cH}^{i_5 i_6)} =0~.
\label{3.2}
\eea
This algebraic constraint is one of the several nonlinear constraints, which 
define the irreducible linear Goldstino superfield ${\mathbb H}^{ij}$ introduced in \cite{KMcAT-M}. 
As will be shown in Appendix \ref{App2}, the  cubic constraint \eqref{3.2}
expresses the $\sSU(2)$ triplet of physical scalars, $\cH^{ij}|_{\q=0}$,
in terms of the other component fields of $\cH^{ij}$. Thus the field content of 
 $\cH^{ij}$ is as follows: two Goldstini, a gauge two-form, and a complex nowhere vanishing auxiliary scalar. 
 As for the Goldstino superfield ${\mathbb H}^{ij}$, 
its only independent component fields are the Goldstini, since 
the additional nonlinear constraints, which  ${\mathbb H}^{ij}$ obeys, 
express the gauge two-form and the auxiliary fields in terms of the Goldstone
fermions \cite{KMcAT-M}. 

To describe the dynamics of $\cN=2$ supergravity coupled to $\cH^{ij}$
we choose an action of the form
\bea
S= S_{\rm SUGRA}  +S_\cH~, 
\eea
where the supergravity action is given by \eqref{6.11}. 
The action $S_\cH$ for the Goldstino superfield has, probably, the simplest 
form within  the projective-superspace formulation for $\cN=2$ supergravity 
\cite{KLRT-M2,KLRT-M1}. Here we refer the reader to \cite{KLRT-M2,KLRT-M1}
for the technical details 
of that formulation, and we simply give the projective superfield Lagrangian
corresponding to $S_\cH$. Using the modern projective-superspace notation
\cite{Kuzenko:2010bd}, the Lagrangian is
\bea
\cL^{(2)}_\cH = -\hf \frac{ \cH^{(2)} \cH^{(2)} }{G^{(2)}}
+\x V \cH^{(2)}~,
\label{3.4}
\eea
with $\x$ being a coupling constant. Here we have denoted 
$\cH^{(2)} = \cH_{ij} v^i v^j$, $G^{(2)} = G_{ij} v^i v^j$, 
where $ v^i \in {\mathbb C}^2 \setminus \{0\}$ denotes the homogeneous 
coordinates for ${\mathbb C}P^1$. Finally, the superfield $V(v^i)$  in \eqref{3.4} is the 
tropical prepotential for the compensating vector multiplet, in particular it is a holomorphic homogeneous function of 
$v^i$ of degree zero. 
 
 The action $S_\cH$ can also be written, in a reasonably compact form, 
 in the conventional curved superspace using the techniques developed in \cite{BK11}. 
It is 
\bea
 S_\cH = -\hf  \int \rd^4 x {\rm d}^4\q \, \cE \, 
\J {\mathbb W}_2  + {\rm c.c.}
     + \x\int \rd^4 x {\rm d}^4\q  \rd^4\bar\theta\,
     E \, \cH^{ij}V_{ij}~,~~~~~~
\label{3.5}
\eea
where ${\mathbb W}_2 $ denotes the reduced chiral superfield 
\cite{BK11} 
\bea
{\mathbb W}_2 &=& - \frac{G}{16} (\bar \cD_{ij} + 4 \bar S_{ij}) \cR_2^{ij}~, 
\quad
\cR_2^{ij} = \frac{1}{ {G}^{4}} \left(\delta^{ij}_{kl} - \frac{1}{2 {G}^2} {G}^{ij} {G}_{kl} \right)
     \cH^{(kl} \cH^{m n)} {G}_{mn}~.
\eea

The action \eqref{3.5} can be generalised to include higher derivative 
terms that can be constructed using the techniques developed in \cite{BK11}.

%%%%%%%%%%%%%%%%%%%%%%%%%%%%%%%%%
%%%%%%%%%%%%%%%%%%%%%%%%%%%%%%%%

\section{Nilpotent twisted chiral superfields}

In this section we restrict our attention to
the case of $\cN=2$ Poincar\'e supersymmetry and introduce 
new nilpotent superfields on Minkowski superspace
${\mathbb M}^{4|8} $ parametrised by Cartesian coordinates 
$z^A = (x^a, \q^\a_i , \bar \q^i_\ad)$, where $\bar \q^{\ad i} $ is 
the complex conjugate of $\q^\a_i$, with  $i=\1,\2$.
To start with, we recall some salient features of the so-called projective 
supermultiplets that live in the generalised  $\cN=2$  superspace
${\mathbb M}^{4|8} \times {\mathbb C}P^1$ \cite{KLR,LR1,LR2,G-RRWLvU},
see \cite{Kuzenko:2010bd} for a pedagogical review.\footnote{The superspace 
${\mathbb M}^{4|8} \times {\mathbb C}P^1$ was introduced for the first time  by Rosly \cite{Rosly}.
The same superspace is at the heart of the harmonic  \cite{GIKOS,GIOS}
and projective \cite{KLR,LR1,LR2} superspace approaches.} 
As usual,  the notation $D_A=(\pa_a , D^\a_i, \bar D^i_\ad) $ is used for the superspace covariant derivatives. We 
denote by $\z$ the inhomogeneous complex coordinate for ${\mathbb C}P^1$.

An $\cN=2$ superfield $\X(z,\z)$
of the general form
\bea
\Xi (z, \z) = \sum_{n=-\infty}^{+\infty} \Xi_n (z) \z^n
\label{generalprjsup}
\eea
is called projective if it satisfies the constraints
\bsubeq \label{pc1}
\bea
\nabla_\a (\z) \Xi (\z) &=& 0 ~,\qquad  \de_\a(\z)=\z D_\a^\1 -D_\a^\2~, \\
{\bar \nabla}_\ad (\z)\Xi (\z) &=& 0~, \qquad
\deb^\ad(\z)=\DB^\ad_\1 +\z\DB^\ad_\2~.
\eea
\esubeq
These constraints are equivalent to the following differential conditions
\bea\label{projective-constr-2}
&D_\a^\2\X_n
=
D_\a^\1 \X_{n-1}
~,~~~~~~
\DB^\ad_\2 \X_n
=
-\DB^\ad_\1 \X_{n+1}~,
\eea
which imply
\bea
(D^\2)^2\X_n
=
(D^\1)^2\X_{n-2}
~,~~~~~~
(\DB_\2)^2\X_n
=
(\DB_\1)^2\X_{n+2}
~.
\label{4.4}
\eea

Let us now consider a  projective superfield $\U(\z)$ whose Laurent series is
bounded below. Without loss of generality, it can be represented by a Taylor series  
\bea
\label{Xi}
\U(\z)&=&\sum_{n=0}^{+\infty}\U_n\z^n~.
\eea
Then the constraints \eqref{projective-constr-2} tell us that 
the lowest component of $\U(\z)$, $\U_0$, satisfies
chiral and antichiral constraints 
\bea
\DB^\ad_\1\U_0=0~, \qquad D_\a^\2\U_0=0~,
\label{4.6}
\eea
while the next-to-lowest component $\U_1$ obeys linear constraints 
\bea
(\DB_\1)^2\U_1 = 0~, \qquad (D^\2)^2\U_1 = 0~.
\eea
Making use of the constraints \eqref{projective-constr-2} and \eqref{4.4} also gives
\begin{subequations}
\bea
\DB^\ad_\2\U_0&=& - \DB^\ad_\1\U_1~, \\
(\DB_\2)^2 \U_0&=& ( \DB_\1)^2\U_2~.
\eea
\end{subequations}

Constraints of the type \eqref{4.6} were considered for the first time 
thirty five years ago by Galperin, Ivanov and Ogievetsky 
\cite{GIO} in the context of the Fayet-Sohnius hypermultiplet
\cite{Fayet,Sohnius}. 
Recently they have been re-discovered, 
without any reference to \cite{GIO}  and the projective-superspace literature, 
in \cite{ADM-2017}. These authors introduced a ring of $\cN=2$ superfields
$\O$ constrained by 
\bea
\DB^\ad_\1\O=0~, \qquad D_\a^\2\O=0~.
\label{4.9}
\eea
Such superfields were called ``chiral-antichiral''  in \cite{ADM-2017}.
Instead we will call them ``twisted chiral superfields''
by analogy with the two-dimensional terminology introduced in 
\cite{GHR}. 
The most general twisted chiral superfield has the form 
\bea
\O (x, \q_i , \bar \q^j) 
=\re^{-\ri(
\q_\1\s^a\bar{\q}^\1
-\q_\2\s^a\bar{\q}^\2)\pa_a}
\hat \O (x, \q_\1,\bar{\q}^\2)~,
\eea
where $\hat \O (x, \q^\a_\1,\bar{\q}_\ad^\2)$ is an arbitrary function of the four Grassmann variables 
$\q^\a_\1 $ and $\bar{\q}_\ad^\2$. 
We will show that every twisted chiral superfield $\O$ is the lowest component 
of a projective superfield $\U(\z)$.

Given a projective superfield $\X(\z) $, the constraints \eqref{projective-constr-2} 
imply  that the dependence of the component superfields 
$\X_n$ 
on $\q^\a_{\2}$ and ${\bar \q}^{\2}_{ \ad}$ 
is uniquely determined in terms 
of their dependence on $\q^\a_{\1}\equiv \q^\a$
and ${\bar \q}^{\1}_{ \ad}\equiv {\bar \q}_{ \ad}$.  In other words, 
the projective superfield depends effectively 
on half the Grassmann variables which can be chosen
to be the spinor coordinates of the  $\cN=1$ Minkowski superspace
${\mathbb M}^{4|4} $
parametrised by the coordinates $(x^a, \q^\a, \bar \q_\ad)$.
We introduce the spinor covariant derivatives for ${\mathbb M}^{4|4} $, 
$D_\a := D_\a^\1$ and $\bar D^\ad := \bar D^\ad_\1$.
Associated with every  $\cN=2$ superfield $U$ is 
its $\cN=1$ bar-projection $U|:=U|_{\q_\2=\bar{\q}^\2=0}$, which is
an $\cN=1$ superfield.
As we have mentioned, all information about the projective 
multiplet $\X(\z) $ is encoded in its bar-projection $\X(\z) |$.
In particular, associated with the projective multiplet \eqref{Xi} is the 
following family of $\cN=1$ superfields
\bea
\U(\z)| = \f + \z \G +\sum_{n=2}^{+\infty}\U_n |\z^n~, 
\qquad \bar D^\ad \f =0~, \qquad \bar D^2 \G =0~.
\eea
The explicit structure of the $\cN=1$ superfields $\U_n|$, with $n=2,3,\dots$,
depends on the original projective multiplet.

Let us forget for a moment about the projective multiplets and consider 
a twisted chiral superfield $\O$. All information about $\O$ is encoded
in the three $\cN=1$  superfields
\bea\label{components-twisted}
\f:=\O|
~,\qquad
\U^\ad:=
\frac12 \DB^\ad_\2\O|
~,\qquad
\Psi:=-\frac{1}{4}(\DB_2)^2\O|
~,
\eea
all of which are  chiral,
\bea
\DB^\ad\f=0 ~, \qquad 
\DB^\ad\U^\bd=0~, \qquad \DB^\ad\Psi=0~,
\eea
by construction. The chirality of $\U^\ad$ implies
$\U^\ad = -\frac{1}{4} \bar D^2 \L^\ad = \hf \bar D^\ad \bar D_\bd \L^\bd
\equiv -\ft12\DB^\ad \G $.
Thus, there exist $\cN=1$ superfields $\G:=\U_1|$ and $U:=\U_2|$, 
of which $\G$  obeys the linear constraint  $\DB^2\G=0$, 
such that 
\bea
\U^\ad
=-\frac12\DB^\ad \G~, \qquad \Psi=-\frac{1}{4}\DB^2U~.
\eea
Thus we have demonstrated  that every twisted chiral superfield is the lowest 
component of a projective superfield.
In what follows, we do not indicate explicitly the bar-projection.

We now turn to reviewing the structure of supersymmetric actions constructed in terms of the projective multiplets. 
As is well known, associated with every projective multiplet
\eqref{generalprjsup}
 is its smile-conjugate 
\bea
\breve{\X}(\z):=\sum_{n=-\infty}^{+\infty} {(-1)^n}\z^{n}\bar{\X}_{-n}~,
\eea
which is also a projective multiplet.
If the theory is formulated in terms of a projective multiplet 
$\U (\z)$ and its smile-conjugate $\breve{\U}(\z) $, the dynamics is described
with the aid of  a Lagrangian $ \cL(\z) \equiv
\cL (\U (\z),\breve{\U}(\z) , \z)$, which is a projective multiplet. 
Using  this Lagrangian, one can construct a manifestly $\cN=2$ 
supersymmetric action, see \cite{Kuzenko:2010bd} for a pedagogical review. 
As explained in  \cite{Kuzenko:2010bd}, the manifestly $\cN=2$ 
supersymmetric action can be recast in two different but equivalent forms:
\begin{subequations}\label{AAA}
\bea
S&=& \frac{1}{ 16}\oint  \frac{\z \rd\z }{ 2\pi\ri}
\int\rd^4 x\,
({D}^{\1})^2({\bar D}_{\2})^2\cL(\z)
=
\oint  \frac{\z\rd\z }{ 2\pi\ri} \int\rd^4x\rd^2\q_\1\rd^2\bar{\q}^\2 \,
\cL(\z)
\label{ac2}
\\
 &=& \frac{1}{ 16}\oint  \frac{\rd\z }{ 2\pi\ri \z}
\int\rd^4 x\,\z\,
({D}^{\1})^2({\bar D}_{\1})^2\cL(\z)
=\oint \frac{{\rm d}\z}{2\pi\ri\z}
\int\rd^4x\rd^2{\q}\rd^2{\qb}\,  \cL ( \z) 
~,
\label{action}
\eea
\end{subequations}
of which the latter is used in most applications.

We now consider an important example of applying the action principles \eqref{ac2}
and \eqref{action}.
As an extension of the construction given in  \cite{GHK}, 
we choose  $\cL ( \z) = -  F(\U (\z) ) \z^{-2}$, 
with  $F(z)$ being a holomorphic function of one argument, and consider the action
\bea
S =-   \oint_C \frac{{\rm d}\z}{2\pi{\rm i}\z}
\int\rd^4x\rd^2{\q}\rd^2{\qb}\, \frac{  F(\U (\z))  }{\z^2} + {\rm c.c.} ~,
\label{tensoraction}
\eea
where $C$ is a contour around the origin.
Performing the contour integral 
gives
\bea
S&=&
-\int\rd^4x\rd^2{\q}\rd^2{\qb}\big\{
F' (\f) U
+ \ft12 F'' (\f) \, \G^2
\big\}
+{\rm c.c.}
\non\\
&=&
\int\rd^4x\rd^2{\q}\big\{
F''(\f)\U_\ad\U^\ad
-F'(\f)\Psi
\big\}
+{\rm c.c.}
\label{4.18}
\eea
On the other hand, making use of  \eqref{ac2}
leads to the action 
\bea
S=-\int\rd^4x\rd^2\q_\1\rd^2\bar{\q}^\2 \,F(\U_0)
+{\rm c.c.}~, 
\label{tc-action}
\eea
which is an example of the  twisted chiral supersymmetric action
\bea
S_{\rm TC} =\int\rd^4x\rd^2\q_\1\rd^2\bar{\q}^\2 \,\cL_{\rm TC} ~, 
\qquad \DB^\ad_\1\cL_{\rm TC} =0~, \quad D_\a^\2 \cL_{\rm TC} =0~.
\label{4.20}
\eea

The $\cN=2$ supersymmetric theory introduced in \cite{GHK} made use 
of a short projective multiplet 
\bea
H(\z)=H_0+\z H_1-\z^2\bar{H}_0
~,~~~~~~ 
\overline{H_1}=H_1~,
\label{4.21}
\eea
which is known under three different names: 
(i) real $\cO(2)$ multiplet; (ii)  linear multiplet; and (iii) tensor multiplet.
Its $\cN=1$ components include 
  a chiral scalar $\f:=H_0|$
   and a real linear superfield
$G:=H_1|=\overline{G}$,
$D^2G=\DB^2 G=0$.
The $\cN=2$ superfield $H_0$ in $H(\z)$ will be called  a 
{\it short twisted chiral superfield}.
Its $\cN=1$ components 
in \eqref{components-twisted} satisfy
\bea
\U^\ad=-\frac12\DB^\ad G~, \qquad
\Psi=\frac14\DB^2\bar{\f}~.
\eea
The action \eqref{4.18} corresponding to the  $\cO(2)$ multiplet \eqref{4.21} 
 reads \cite{GHK} 
\bea
S=\int\rd^4x\rd^2{\q}\rd^2{\qb}\,
\Big\{
\bar \f F' (\f) +\f {\bar F}' (\bar \f) 
-\hf  \Big( F'' (\f) +\bar F'' (\bar \f) \Big) G^2
\Big\}~,
\label{4.23}
\eea
which is a special case of the general models for self-interacting $\cN=2$ 
tensor multiplets \cite{LR83}.
Dualising the linear superfield $G$ in \eqref{4.23} into  a chiral scalar, 
one ends up with a hyperk\"ahler sigma model. 
The generalisation of  \eqref{4.23} to the case of several $\cN=2$ tensor multiplets, which was given in \cite{GHK}, 
provides a superspace derivation of the rigid $c$-map 
construction \cite{Cecotti:1988qn}.

As was shown in \cite{ADM-2017},  there exists  a simple deformation of the short twisted chiral superfield 
that can be used to derive 
the tensor Goldstone multiplet  for 
partial $\cN=2\to\cN=1$ supersymmetry breaking \cite{BG2} 
from $\cN=2$ superfields. Such a framework 
is actually closely related to the earlier work of \cite{RT,G-RPR}.
To describe partial $\cN=2\to\cN=1$  breaking of supersymmetry, 
the authors of   \cite{RT,G-RPR}
deformed the real $\cO(2)$ multiplet $H(\z)$ 
to a complex $\cO(2)$ multiplet ${\bm H}(\z)$ given by 
\bea
{\bm H}(\z)
=
{\bm H}_0+\z {\bm H}_1+\z^2{\bm H}_2
:=
\widehat{H}(\z)
+m\Big(
(\qb^\2)^2
-\z(\qb^\1\qb^\2)
+\z^2(\qb^\1)^2
\Big)
~.
\eea
Here $\widehat{H}(\z)$ has the functional form \eqref{4.21} and  obeys 
the analyticity conditions
\bea
\nabla_\a (\z) \widehat{H}(\z) = 0 ~,\qquad  
{\bar \nabla}_\ad (\z) \widehat{H}(\z)~, 
\eea
but it does not transform as an $\cN=2$ superfield, unlike  ${\bm H}(\z)$.
The {\it deformed short twisted chiral multiplet} ${\bm H}_0$ has the properties
\bea
{\bm H}_0|=\f
~,~~~~
\U^\ad:=
\frac12 \DB^\ad_\2{\bm H}_0|
=-\frac12\DB^\ad G
~,~~~~
\Psi:=-\frac{1}{4}(\DB_2)^2{\bm H}_0
=\frac14\DB^2\bar{\f}
+m
~,~~~
\label{4.26}
\eea
which coincide with those of the deformed chiral-antichiral multiplet considered in \cite{ADM-2017}.
The mass parameter $m$ \eqref{4.26}
plays a role similar to the deformation parameter of the 
deformed reduced chiral superfield in the flat superspace limit.
The presence of the deformation parameter $m$ modifies the second supersymmetry transformation:
\bsubeq\label{tc-susy2}
\bea
\d \f&=&-2\bar{\e}^\2_\ad \U^\ad 
~,
\\
\d \U^\ad
&=&
\frac12 m\bar{\e}^{\ad\2}
+\frac14\bar{\e}^{\ad\2} \DB^2\bar\f
-\ri\e_\2^\a \pa_\a{}^\ad \f
~.
\eea
\esubeq
The action 
\eqref{4.20}
with a  Lagrangian $\cL_{\rm TC}  =- F({\bm H}_0)$
takes,  upon reduction to 
$\cN=1$ superspace, the following form:
\bea
S &=& 
\int{\rd}^{4}x\rd^2{\q}\rd^2{\qb}\,W(\f) \bar\f
+\int{\rm d}^{4}x\rd^2{\q}\,\big\{ 
W'(\f)\U_\ad\U^\ad
+mW(\f)
\big\}
+{\rm c.c. }
~,~~~~~
\eea
where $W(\f):=F'({\bm H}_0)|$.

To describe $\cN=2\to\cN=1$ supersymmetry breaking, 
it remains to impose the quadratic nilpotency condition
\cite{ADM-2017} 
\bea
{\bm H}_0{}^2=0
~,
\label{dtc2}
\eea
in agreement with the earlier results of \cite{RT,G-RPR}.
It terms of  $\cN=1$ superfields, 
this constraint is equivalent to 
\bea
\f^2=0
~,~~~
\f\Upsilon^\ad=0
~,~~~
\Big(m
+\frac14\DB^2\bar{\f}
\Big)\f
=
\U_\ad\U^\ad
~,
\label{dtc2-2}
\eea
which are exactly the Bagger-Galperin constraints
\cite{BG2}.

Instead of imposing  the constraint \eqref{dtc2}, 
we now consider a cubic nilpotency condition
\bea
{\bm H}_0{}^3=0
~.
\label{dtc3}
\eea
Upon reduction to $\cN=1$ superfields, it  implies
\bea
\f^3=0
~,~~~
\f^2\Upsilon^\ad=0
~,~~~
\Big(m
+\frac14\DB^2\bar{\f}
\Big)\f^2
=
\f\U_\ad\U^\ad
~.
\label{dtc3-2}
\eea
These constraints were introduced in \cite{DFS}.
Our analysis derives them in the full $\cN=2$ superspace in terms of a deformed 
short  twisted chiral  Goldstone multiplet.
As discussed in \cite{DFS} the solution of \eqref{dtc3-2} mimics the case of the deformed reduced chiral 
Goldstone multiplet subject to a cubic nilpotent constraint. 
The solution includes two branches: i) one which is identical to the $\cN=2\to\cN=1$ 
supersymmetry breaking case
solving \eqref{dtc2-2};
and ii) 
one that completely breaks supersymmetry in general and determines
${\bm H}_0$ in terms of the following physical degrees of freedom:
a scalar, two Goldstini,  and a gauge two-form  \cite{DFS}. 
Note that for the first branch to exist it is necessary to have the mass parameter 
to be non-vanishing, $m\ne 0$.
This feature distinguishes the present deformed short twisted chiral model from a nilpotent linear multiplet.

In this paper, we did not describe the component structure of the supergravity-matter theories proposed. 
 These theories can be reduced to components using the results of \cite{BN}.
\\
%%%%%%%%%%%%%%%%%%%%%%%%%%%%%%
%%%%%%%%%%%%%%%%%%%%%%%%

\noindent
{\bf Acknowledgements:}\\
We are grateful to Joseph Novak for comments on the manuscript. 
The work of SMK is supported
in part by the Australian Research Council, project No.\,DP160103633.
The work of GT-M was supported by the Interuniversity
Attraction Poles Programme initiated by the Belgian Science Policy (P7/37).

%%%%%%%%%%%%%%%%%%%%%%%%%%%%%%%%%%%%%%%%%%%%%%%%%%%%%%

\appendix 

\section{Reduced chiral and linear multiplets}\label{App1}

It is well known that the field strength of an Abelian vector multiplet is a reduced chiral superfield \cite{GSW}.
In curved superspace, it is a
covariantly chiral superfield $W$,  
\begin{subequations}
\bea
\cDB^\ad_i W&=& 0~,
\eea
 subject to the Bianchi identity \cite{GSW,Howe}
 \bea
\big(\cD^{ij}+4S^{ij}\big) W&=&
\big(\cDB^{ij} +4\bar{S}^{ij}\big)\bar{W} ~.
\eea
\end{subequations}

We recall that the $\cN=2$ tensor multiplet is  described in curved superspace by
its gauge-invariant field strength $G^{ij}$  which is 
a linear multiplet. The latter is 
defined to be a  real ${\sSU}(2)$ triplet (that is, 
$G^{ij}=G^{ji}$ and ${\bar G}_{ij}:=\overline{G^{ij}} = G_{ij}$)
subject to the covariant constraints  \cite{BS,SSW}
\bea
\cD^{(i}_\a G^{jk)} =  {\bar \cD}^{(i}_\ad G^{jk)} = 0~.
\eea
These constraints are solved in terms of a chiral
prepotential $\Psi$ \cite{HST,GS82,Siegel83,Muller86} via
\begin{align}
\label{eq_Gprepotential}
G^{ij} = \frac{1}{4}\big( \cD^{ij} +4{S}^{ij}\big) \Psi
+\frac{1}{4}\big( \cDB^{ij} +4\bar{S}^{ij}\big){\bar \Psi}~, \qquad
{\bar \cD}^i_\ad \J=0~,
\end{align}
which is invariant under Abelian gauge transformations
\begin{subequations}\label{TMGT}
\bea
\d_\L \Psi = \ri \Lambda~,
\eea
with the gauge parameter $\Lambda$ being a reduced chiral superfield,
\bea
\bar \cD^i_\ad \L &=&0~, \qquad
\big(\cD^{ij}+4S^{ij}\big)\L
- \big(\cDB^{ij}+ 4\bar{S}^{ij}\big)\bar{\L} = 0~.
\eea
\end{subequations}

The constraints on $\L$ can be solved in terms of 
the Mezincescu prepotential \cite{Mezincescu} (see also \cite{HST}),  $V_{ij}=V_{ji}$,
which is an unconstrained real $\sSU(2)$ triplet. 
The curved-superspace solution is \cite{BK11}
\begin{align}
\L = \frac{1}{4}\bar\Delta \big({\cD}^{ij} + 4 S^{ij}\big) V_{ij}~.
\end{align}
Here   $\bar{\D}$ denotes the chiral projection operator \cite{Muller}
\bea
\bar{\D}
&=&\frac{1}{96} \Big(\big(\cDB^{ij}+16\bar{S}^{ij}\big)\cDB_{ij}
-\big(\cDB^{\ad\bd}-16\bar{Y}^{\ad\bd}\big)\cDB_{\ad\bd} \Big)
\non\\
&=&\frac{1}{96} \Big(\cDB_{ij}\big(\cDB^{ij}+16\bar{S}^{ij}\big)
-\cDB_{\ad\bd}\big(\cDB^{\ad\bd}-16\bar{Y}^{\ad\bd}\big) \Big)~,
\label{chiral-pr}
\eea
with $\cDB^{\ad\bd}:=\cDB^{(\ad}_k\cDB^{\bd)k}$.
Its main properties can be formulated using 
a super-Weyl inert scalar $U$. It holds that
\begin{subequations} 
\bea
{\bar \cD}^{\ad}_i \bar{\D} U &=&0~, \\
\d_\s U = 0 \quad \Longrightarrow \quad 
\d_\s \bar \D U &=& 2\s \bar \D U~,  \label{2.5b}\\
\int \rd^4 x {\rm d}^4\q{\rm d}^4{\bar \q}\,E\, U
&=& \int {\rm d}^4x {\rm d}^4 \q \, \cE \, \bar{\D} U ~,
\label{chiralproj1} 
\eea
\end{subequations}
where $\s$ is the real  unconstrained parameter of a super-Weyl transformation
\cite{Howe,KLRT-M2}. The detailed derivation of \eqref{chiralproj1} is given in 
\cite{KT-M09}

%%%%%%%%%%%%%%%%%%%%%%
%%%%%%%%%%%%%%%%%%%%%%%%

\section{Solving the nilpotency condition \eqref{3.2}}\label{App2}

In this appendix we show how to solve the nilpotency condition \eqref{3.2}
in Minkowski superspace.  We make use of the harmonic superspace techniques
\cite{GIKOS,GIOS}.

Associated with the linear superfield $\cH^{ij} (z)$ constrained by \eqref{3.1}
is the harmonic superfield $\cH^{++} (z,u) := \cH^{ij} (z)u^+_i u^+_j$
which is analytic and short:
\begin{subequations}
\bea
D^+_\a \cH^{++}&=&0~, \quad \bar D^+_\ad \cH^{++}=0~, \\ 
D^{++} \cH^{++} &=&0~.
\eea
\end{subequations}
The analyticity constraints mean that $\cH^{++}$ lives
on the  analytic subspace of the harmonic superspace parametrised by 
$\zeta_\rA\equiv\{x^m_\rA,\theta^{+\alpha},{\bar\theta}^+_{\dot\alpha}\}$ 
and $u^\pm_i$. Here the variables 
\bea
x_\rA^m = x^m - 2{\rm i}\theta^{(i}\sigma^m
{\bar \theta}^{j)}u^+_iu^-_j~, \qquad 
\theta^\pm_\alpha=u^\pm_i\theta^i_\alpha~, \qquad {\bar\theta}^\pm_{\dot\alpha}
=u^\pm_i{\bar \theta}^i_{\dot\alpha}
\eea
 correspond to the analytic basis of 
 the harmonic superspace. 

In the analytic basis, the general expression for $\cH^{++}$ 
was given in \cite{GIOS}. It is 
\bea
\cH^{++}(\z_\rA,u)
&=&
h^{ij}(x_\rA)u^+_iu^+_j
+2\big{[}
\q^{+\a}\psi_\a^i(x_\rA)
-\qb_\ad^{+}\bar\psi^{\ad i}(x_\rA)
\big{]}u_i^+
\non\\
&&
+(\q^+)^2M(x_\rA)
+(\qb^+)^2\bar{M}(x_\rA)
\non\\
&&
+2\ri \q^+\s^m\qb^+
\big{[}
V_m(x_\rA)
+\pa_m h^{ij}(x_\rA)
u^+_iu^-_j
\big{]}
\non\\
&&
+2\ri\big{[}
(\qb^+)^2\q^{+\a}\pa_{\a\ad}\bar\psi^{\ad i}(x_\rA)
+(\q^+)^2\qb_\ad^{+}\pa^{\a\ad}\psi_\a^{i}(x_\rA)
\big{]}
u^-_i
\non\\
&&
+(\q^+)^4\Box h^{ij}(x_\rA)u^-_iu^-_j~. 
\eea
Here 
$h_{ij}=\overline{ h^{ij} }$, 
${\bar \psi}_{\ad i} \equiv \overline{\psi_\a^i }$, 
and $V^m$ is a real conserved vector, 
\bea
\pa_m V^m =0~,
\eea
which allows us to interpret $V^m$ as the Hodge dual of the field strength of a gauge two-form.
Here one should keep in mind that 
the operator $D^{++} = u^{+i} {\pa}/{\pa u^{-i}} $ in the analytic basis takes the form 
\bea
D^{++}_\rA = u^{+i} \frac{\pa}{\pa u^{-i}} -2\ri \q^+ \s^m \bar \q^+ 
\frac{\pa }{\pa x^m_\rA} +\dots~,
\eea
where the ellipsis denotes two additional terms which do not contribute 
when acting on analytic superfields.

In the harmonic superspace setting, the nilpotency condition takes the form
\bea
(\cH^{++})^3 =0~.
\label{B.5}
\eea
At the component level, 
this condition is equivalent to the following equations
\begin{subequations}\label{B.6}
\bea
0&=&(h^{++})^3~,  
\\
0&=&(h^{++})^2\j_\a^{+} ~,
 \\
0&=& h^{++}\big(h^{++}M  -2(\j^+)^2\big)~, \label{B.6c}\\
0&=& h^{++}\big(\,\ri h^{++} (V_m +\pa_m h^{+-})
+2  \j^+ \s_m \bar \j^+\big)
~,
\\
0&=&
h^{++}\Big(M\bar\psi_\ad^{+}
+\ri (V_m+\pa_mh^{+-})(\psi^{+}\s^m)_{\ad} 
+\frac{\ri}{2} h^{++}(\pa_m\psi^{-}\s^m)_{\ad}
\Big)
-(\psi^{+})^2\bar\psi_\ad^{+}
~,
~~~
\\
0&=&
h^{++}\Big(
M\bar{M}
+(V_{m}+\pa_mh^{+-})^2
-2\ri \psi^{+}\s^m{\partial}_m\bar\psi^{-}
-2\ri \pa_m\psi^{-}\s^m\bar\psi^{+}
+\hf h^{++}\Box h^{--}
\Big)
\non\\
&&
-(\psi^{+})^2\bar{M}
-(\bar\psi^{+})^2M
-2\ri (V_m+\pa_mh^{+-})\psi^+\s^m\bar\psi^{+}
~,
\eea
\end{subequations}
where we have introduced $h^{\pm\pm} := h^{ij} u^\pm_i u^\pm_j$ 
and
$\j^\pm_\a := \j^i_\a u^\pm_i$,
The equations \eqref{B.6} are solved by
%%%%%%
\bea
h^{ij}
&=&
\frac{
\psi^{(i}\psi^{j)}\bar{M}
+\bar\psi^{(i}\bar\psi^{j)}M
+2\ri \psi^{(i}\s_m\bar\psi^{j)}V^{m}}{M\bar{M}+V^{n}V_{n}} +\dots ~,
\eea
where the ellipsis denotes all terms with derivatives of the fields.
It  is  assumed that the complex auxiliary field $M$ is nowhere vanishing, 
$M \neq 0$, 
and the allowed values of the field strength $V^m$ are restricted by 
\bea
M\bar{M}+V^{n}V_{n} \neq 0~.
\eea
We will present the complete solution elsewhere. However it should be pointed 
out that $h^{ij} $ vanishes if the Goldstini are switched off, 
\bea
\j^i_\a =0 \quad \Longrightarrow \quad h^{ij}=0~.
\eea 
Indeed, in the case  $\j^i_\a =0$ eq. \eqref{B.6c} reduces to $h^{++}h^{++}M =0$.
This implies $h^{ij}=0$  if the components of $h^{ij}$ are ordinary complex numbers, 
as a consequence of the identity $h^{ij} = \ri q^{(i} \bar q^{j)}$, 
for some $\sSU(2)$ spinor $q^i$ and its conjugate $\bar q_i = \overline{q^i}$.

%%%%%%%%%%%%%%%%%%%%%%%%%%%%%%%%%%%%%%%%%%%%%%%%%%%%%%%%%%

\begin{footnotesize}

\end{footnotesize}


\begin{thebibliography}{66}

 
 \bibitem{ADFS}
I.~Antoniadis, E.~Dudas, S.~Ferrara and A.~Sagnotti,
  ``The Volkov-Akulov-Starobinsky supergravity,''
  Phys.\ Lett.\ B {\bf 733}, 32 (2014) 
  [arXiv:1403.3269 [hep-th]].
 


\bibitem{DFKS} 
  E.~Dudas, S.~Ferrara, A.~Kehagias and A.~Sagnotti,
  ``Properties of nilpotent supergravity,''
  JHEP {\bf 1509}, 217 (2015)
  [arXiv:1507.07842 [hep-th]].


 \bibitem{BFKVP} 
  E.~A.~Bergshoeff, D.~Z.~Freedman, R.~Kallosh and A.~Van Proeyen,
  ``Pure de Sitter supergravity,''
  Phys.\ Rev.\ D {\bf 92}, no. 8, 085040 (2015)
  Erratum: [Phys.\ Rev.\ D {\bf 93}, no. 6, 069901 (2016)]
  [arXiv:1507.08264 [hep-th]]. 
  
\bibitem{HY} 
  F.~Hasegawa and Y.~Yamada,
  ``Component action of nilpotent multiplet coupled to matter in 4 dimensional $ \mathcal{N}=1 $ supergravity,''
  JHEP {\bf 1510}, 106 (2015)
  [arXiv:1507.08619 [hep-th]].  
  
\bibitem{K15} 
S.~M.~Kuzenko,
``Complex linear Goldstino superfield and supergravity,''
JHEP {\bf 1510}, 006 (2015)
[arXiv:1508.03190 [hep-th]].

\bibitem{BMST} 
I.~Bandos, L.~Martucci, D.~Sorokin and M.~Tonin,
``Brane induced supersymmetry breaking and de Sitter supergravity,''
JHEP {\bf 1602}, 080 (2016)
[arXiv:1511.03024 [hep-th]].



\bibitem{FKRR} 
  F.~Farakos, A.~Kehagias, D.~Racco and A.~Riotto,
  ``Scanning of the supersymmetry breaking scale and the gravitino mass in supergravity,''
  JHEP {\bf 1606}, 120 (2016)
  [arXiv:1605.07631 [hep-th]].

\bibitem{BHKMS} 
I.~Bandos, M.~Heller, S.~M.~Kuzenko, L.~Martucci and D.~Sorokin,
  ``The Goldstino brane, the constrained superfields and matter in $ \mathcal{N}=1 $ supergravity,''  
  JHEP {\bf 1611}, 109 (2016)
  [arXiv:1608.05908 [hep-th]].


\bibitem{BK17} 
  E.~I.~Buchbinder and S.~M.~Kuzenko,
  ``Three-form multiplet and supersymmetry breaking,''
  arXiv:1705.07700 [hep-th].



\bibitem{KMcAT-M} 
  S.~M.~Kuzenko, I.~N.~McArthur and G.~Tartaglino-Mazzucchelli,
  ``Goldstino superfields in N=2 supergravity,''   JHEP {\bf 1705}, 061 (2017)
[arXiv:1702.02423 [hep-th]].




\bibitem{KMcA} 
  S.~M.~Kuzenko and I.~N.~McArthur,
  ``Goldstino superfields for spontaneously broken N=2 supersymmetry,''
  JHEP {\bf 1106}, 133 (2011)
  [arXiv:1105.3001 [hep-th]].
  

\bibitem{CDF} 
N.~Cribiori, G.~Dall'Agata and F.~Farakos,
``Interactions of N goldstini in superspace,''
Phys.\ Rev.\ D {\bf 94}, no. 6, 065019 (2016)
  [arXiv:1607.01277 [hep-th]].


\bibitem{VA}
D.~V.~Volkov and V.~P.~Akulov,
``Possible universal neutrino interaction,''
  {JETP Lett.\  {\bf 16}, 438 (1972)}   
  [Pisma Zh.\ Eksp.\ Teor.\ Fiz.\   {\bf 16},  621 (1972)]; 
  ``Is the neutrino a Goldstone particle?,''
  Phys.\ Lett.\  B {\bf 46}, 109 (1973).

\bibitem{AV}
V.~P. Akulov and D.~V. Volkov, ``Goldstone fields with spin 1/2,''
   Theor. Math. Phys. {\bf 18}, 28 (1974)  28 [Teor. Mat. Fiz. {\bf 18}, 39 (1974)].


  

\bibitem{KT-M15} 
  S.~M.~Kuzenko and G.~Tartaglino-Mazzucchelli,
  ``Nilpotent chiral superfield in N=2 supergravity and partial rigid supersymmetry breaking,''
  JHEP {\bf 1603}, 092 (2016)
  [arXiv:1512.01964 [hep-th]].
  
  


\bibitem{Rocek} 
M.~Ro\v{c}ek,
``Linearizing the Volkov-Akulov model,''
  Phys.\ Rev.\ Lett.\  {\bf 41}, 451 (1978).
  
\bibitem{IK78}
E.~Ivanov and A.~Kapustnikov, ``General relationship between linear and
  nonlinear realisations of supersymmetry,''
  J.\ Phys.\  A {\bfseries 11}   (1978) 2375.



\bibitem{Casalbuoni}
R.~Casalbuoni, S.~De Curtis, D.~Dominici, F.~Feruglio, and R.~Gatto,
  ``{Non-linear realization of supersymmetry algebra from supersymmetric constraint},'' 
   Phys.\ Lett.\  B 
  {\bfseries 220},  569 (1989).


\bibitem{KS}
  Z.~Komargodski and N.~Seiberg,
  ``From linear SUSY to constrained superfields,''
  JHEP {\bf 0909}, 066 (2009)
  \href{http://arxiv.org/abs/0907.2441}{arXiv:0907.2441}.
  


\bibitem{LR79}
U.~Lindstr\"om and M.~Ro\v{c}ek,
``Constrained local superfields,''
Phys.\ Rev.\  D {\bf 19}, 2300 (1979).
  


\bibitem{DFS} 
  E.~Dudas, S.~Ferrara and A.~Sagnotti,
  ``A superfield constraint for $\cN=2 \to \cN=0$ breaking,''
  arXiv:1707.03414 [hep-th].



\bibitem{K-superWeyl} 
  S.~M.~Kuzenko,
  ``Super-Weyl anomalies in N=2 supergravity and (non)local effective actions,''
  JHEP {\bf 1310}, 151 (2013)
  [arXiv:1307.7586 [hep-th]].


\bibitem{deWPV}
B.~de Wit, R.~Philippe and A.~Van Proeyen,
``The improved tensor multiplet in N = 2 supergravity,''
Nucl.\ Phys.\ B {\bf 219}, 143 (1983).


  
  \bibitem{KLRT-M2}
S.~M.~Kuzenko, U.~Lindstr\"om, M.~Ro\v cek and G.~Tartaglino-Mazzucchelli,
``On conformal supergravity and projective superspace,''
JHEP {\bf 0908}, 023 (2009)
[arXiv:0905.0063 [hep-th]].
  

\bibitem{Howe}
P.~S.~Howe, ``Supergravity in superspace,''  
Nucl.\ Phys.\  B {\bf 199}, 309 (1982).




\bibitem{APT}
  I.~Antoniadis, H.~Partouche and T.~R.~Taylor,
``Spontaneous breaking of N=2 global supersymmetry,''
  Phys.\ Lett.\  B {\bf 372}, 83 (1996)
  [arXiv:hep-th/9512006].
  

\bibitem{IZ1} 
  E.~A.~Ivanov and B.~M.~Zupnik,
  ``Modified N=2 supersymmetry and Fayet-Iliopoulos terms,''
  Phys.\ Atom.\ Nucl.\  {\bf 62}, 1043 (1999)
  [Yad.\ Fiz.\  {\bf 62}, 1110 (1999)]
  [hep-th/9710236].


\bibitem{IZ2} 
  E.~Ivanov and B.~Zupnik,
  ``Modifying N=2 supersymmetry via partial breaking,''
  In *Buckow 1997, Theory of elementary particles* 64-69
  [hep-th/9801016].


\bibitem{BIL} 
  D.~Butter, G.~Inverso and I.~Lodato,
  ``Rigid 4D $ \mathcal{N}=2 $ supersymmetric backgrounds and actions,''
  JHEP {\bf 1509}, 088 (2015)
  [arXiv:1505.03500 [hep-th]].


\bibitem{Kuzenko:2016nbu} 
  S.~M.~Kuzenko,
  ``Maximally supersymmetric solutions of $R^2$ supergravity,''
  Phys.\ Rev.\ D {\bf 94}, no. 6, 065014 (2016)
  [arXiv:1606.00654 [hep-th]].

\bibitem{DP}
S.~Deser and R.~Puzalowski, 
``Supersymmetric nonpolynomial vector multiplets and causal propagation,''
J. Phys. A {\bf 13},  2501 ( 1980).

\bibitem{CF}
S.~Cecotti and S.~Ferrara,
``Supersymmetric Born-Infeld Lagrangians,''
Phys.\ Lett.\ B {\bf 187}, 335 (1987).


\bibitem{BG}
J.~Bagger and A.~Galperin,
``A new Goldstone multiplet for partially broken supersymmetry,''
Phys.\ Rev.\ D {\bf 55}, 1091 (1997) [arXiv:hep-th/9608177].



\bibitem{KT2}
S.~M.~Kuzenko and S.~Theisen,
``Supersymmetric duality rotations,''
JHEP {\bf 0003}, 034 (2000)
[arXiv:hep-th/0001068].


\bibitem{KT3}
  S.~M.~Kuzenko and S.~Theisen,
  ``Nonlinear self-duality and supersymmetry,''
Fortsch.\ Phys.\  {\bf 49}, 273 (2001)
[hep-th/0007231].


\bibitem{KMcC}
S.~M.~Kuzenko and S.~A.~McCarthy,
  ``Nonlinear self-duality and supergravity,''
JHEP {\bf 0302}, 038 (2003) [hep-th/0212039].



\bibitem{RT}
M.~Ro\v{c}ek and A.~A.~Tseytlin,
``Partial breaking of global D = 4 supersymmetry, 
constrained  superfields, and 3-brane actions,''
Phys.\ Rev.\ D {\bf 59}, 106001 (1999) 
[arXiv:hep-th/9811232].


 \bibitem{BK11} 
  D.~Butter and S.~M.~Kuzenko,
  ``New higher-derivative couplings in 4D N = 2 supergravity,''
  JHEP {\bf 1103}, 047 (2011)
  [arXiv:1012.5153 [hep-th]].


 
\bibitem{K-08}
S.~M.~Kuzenko,
``On N = 2 supergravity and projective superspace: Dual formulations,''
Nucl.\ Phys.\  B {\bf 810}, 135 (2009)
[arXiv:0807.3381 [hep-th]].
 

 



\bibitem{KLRT-M1}
S.~M.~Kuzenko, U.~Lindstr\"om, M.~Ro\v cek and G.~Tartaglino-Mazzucchelli,
``4D N=2 supergravity and projective superspace,'' 
JHEP {\bf 0809}, 051 (2008) [arXiv:0805.4683].



  \bibitem{KLR}
A. Karlhede, U. Lindstr\"om and M. Ro\v cek,
``Self-interacting tensor multiplets in N=2 superspace,''
Phys.\ Lett.\ B {\bf 147}, 297 (1984).

\bibitem{LR1}
U.~Lindstr\"om and M.~Ro\v{c}ek,
``New hyperk\"ahler  metrics  and new supermultiplets,''
  Commun.\ Math.\ Phys.\  {\bf 115}, 21 (1988).
  
\bibitem{LR2}
U.~Lindstr\"om and M.~Ro\v{c}ek,  
 ``N=2 super Yang-Mills theory in projective superspace,''
Commun.\ Math.\ Phys.\  {\bf 128}, 191 (1990).

\bibitem{G-RRWLvU}
F.~Gonzalez-Rey, M.~Ro\v{c}ek, S.~Wiles, U.~Lindstr\"om and R.~von Unge,
``Feynman rules in N = 2 projective superspace. I: Massless  hypermultiplets,''
Nucl.\ Phys.\  B {\bf 516}, 426 (1998)
[arXiv:hep-th/9710250].


\bibitem{Kuzenko:2010bd} 
  S.~M.~Kuzenko,
  ``Lectures on nonlinear sigma-models in projective superspace,''
  J.\ Phys.\ A {\bf 43}, 443001 (2010)
  [arXiv:1004.0880 [hep-th]].

\bibitem{Rosly}
 A.~A.~Rosly,
``Super Yang-Mills  constraints 
as integrability conditions,'' in Proceedings of the International 
Seminar {\it Group Theoretical 
Methods in Physics} (Zvenigorod, USSR, 1982),
M. A. Markov  (Ed.), 
Nauka, Moscow, 1983, Vol. 1, p. 263 (in Russian);
English translation: in {\it Group Theoretical 
Methods in Physics},'' M. A. Markov, V. I. Man'ko 
and A. E. Shabad  (Eds.), Harwood Academic Publishers, 
London, Vol. 3, 1987, p. 587.

\bibitem{GIKOS}
A.~S.~Galperin, E.~A.~Ivanov, S.~N.~Kalitzin, V.~Ogievetsky, E.~Sokatchev, 
``Unconstrained N=2 matter, Yang-Mills and supergravity theories in harmonic
superspace,''
Class.\ Quant.\ Grav.\  {\bf 1}, 469 (1984).

\bibitem{GIOS}
A.~S.~Galperin, E.~A.~Ivanov, V.~I.~Ogievetsky and E.~S.~Sokatchev,
{\it Harmonic Superspace}, Cambridge University Press,  Cambridge, 2001.




\bibitem{GIO} 
A.~Galperin, E.~Ivanov and V.~Ogievetsky,
  ``Superfield anatomy of the Fayet-Sohnius multiplet,''
  Sov.\ J.\ Nucl.\ Phys.\  {\bf 35}, 458 (1982)
  [Yad.\ Fiz.\  {\bf 35}, 790 (1982)].
  
 

\bibitem{Fayet}
P.~Fayet, ``Fermi-Bose hypersymmetry,''
Nucl.\ Phys.\ B {\bf 113}, 135 (1976).

\bibitem{Sohnius}
  M.~F.~Sohnius,
``Supersymmetry and central charges,''
Nucl.\ Phys.\  B {\bf 138}, 109 (1978).
  
\bibitem{ADM-2017}
  I.~Antoniadis, J.~P.~Derendinger and C.~Markou,
  ``Nonlinear ${\cal N}=2$ global supersymmetry,''
  JHEP {\bf 1706}, 052 (2017)
  [arXiv:1703.08806 [hep-th]].  


\bibitem{GHR}
  S.~J.~Gates Jr., C.~M.~Hull and M.~Ro\v{c}ek,
  ``Twisted multiplets and new supersymmetric nonlinear sigma models,''
Nucl.\ Phys.\  B {\bf 248}, 157 (1984).

  \bibitem{GHK} 
  S.~J.~Gates Jr., T.~H\"ubsch and S.~M.~Kuzenko,
``CNM models, holomorphic functions and projective superspace C maps,''
  Nucl.\ Phys.\ B {\bf 557}, 443 (1999)
  [hep-th/9902211].


\bibitem{LR83}
  U.~Lindstr\"om and M.~Ro\v{c}ek,
  ``Scalar tensor duality and N=1, N=2 nonlinear sigma models,''
  Nucl.\ Phys.\ B {\bf 222}, 285 (1983).

  
  \bibitem{Cecotti:1988qn} 
  S.~Cecotti, S.~Ferrara and L.~Girardello,
  ``Geometry of Type II superstrings and the moduli of superconformal field theories,''
  Int.\ J.\ Mod.\ Phys.\ A {\bf 4}, 2475 (1989).


\bibitem{BG2}
J.~Bagger and A.~Galperin,
``The tensor Goldstone multiplet for partially broken supersymmetry,''
Phys.\ Lett.\  {\bf B412},  296 (1997) 
[hep-th/9707061].


\bibitem{G-RPR} 
  F.~Gonzalez-Rey, I.~Y.~Park and M.~Ro\v{c}ek,
  ``On dual 3-brane actions with partially broken N=2 supersymmetry,''
  Nucl.\ Phys.\ B {\bf 544}, 243 (1999)
  [hep-th/9811130].


\bibitem{BN} 
  D.~Butter and J.~Novak,
  ``Component reduction in N=2 supergravity: the vector, tensor, and vector-tensor multiplets,''
JHEP {\bf 1205}, 115 (2012)
[arXiv:1201.5431 [hep-th]].
  


\bibitem{GSW}
 R.~Grimm, M.~Sohnius and J.~Wess,
  ``Extended supersymmetry and gauge theories,''
Nucl.\ Phys.\  B {\bf 133}, 275 (1978).

  
\bibitem{BS}
P.~Breitenlohner and M.~F.~Sohnius,
``Superfields, auxiliary fields, and tensor calculus for N=2 extended
supergravity,''
Nucl.\ Phys.\  B {\bf 165}, 483 (1980);
``An almost simple off-shell version of SU(2) Poincare supergravity,''
Nucl.\ Phys.\  B {\bf 178}, 151 (1981).



\bibitem{SSW}
  M.~F.~Sohnius, K.~S.~Stelle and P.~C.~West,
 ``Representations of extended supersymmetry,''
in {\it Superspace and Supergravity}, S. W. Hawking and M. Ro\v{c}ek (Eds.), 
Cambridge University Press, Cambridge, 1981, p. 283.  




\bibitem{HST}
P.~S.~Howe, K.~S.~Stelle and P.~K.~Townsend,
``Supercurrents,''  Nucl.\ Phys.\  B {\bf 192}, 332 (1981).




\bibitem{GS82}
S.~J.~Gates Jr. and W.~Siegel,
``Linearized N=2 superfield supergravity,''  Nucl.\ Phys.\  B {\bf 195}, 39 (1982).


\bibitem{Siegel83}
  W.~Siegel,
  ``Off-shell N=2 supersymmetry for the massive scalar multiplet,''
  Phys.\ Lett.\  B {\bf 122}, 361 (1983).

 

\bibitem{Muller86}
  M.~M\"uller,
  ``Chiral actions for minimal N=2 supergravity,''
  Nucl.\ Phys.\  B {\bf 289}, 557 (1987).






\bibitem{Mezincescu}
  L.~Mezincescu,
  ``On the superfield formulation of O(2) supersymmetry,''
  Dubna preprint JINR-P2-12572 (June, 1979).


       
  
 \bibitem{Muller} M. M\"uller, {\it Consistent Classical Supergravity Theories},
(Lecture Notes in Physics, Vol. 336),
Springer, Berlin, 1989. 
 
 

\bibitem{KT-M09}
  S.~M.~Kuzenko and G.~Tartaglino-Mazzucchelli,
 ``Different representations for the action principle in 4D N = 2 supergravity,''
  JHEP {\bf 0904}, 007 (2009) 
  [arXiv:0812.3464 [hep-th]].



\end{thebibliography}
\end{document}